\documentclass{article}

\begin{document}

\title{Kinetics of nucleation for the decreasing rate of the new embryos growth }
\author{Victor Kurasov}
\date{Victor.Kurasov@pobox.spbu.ru}

\begin{abstract}
Kinetics of the first order phase transition is investigated.
In some rather wide spread situations the property of avalanche growth of
the objects of a new phase doesn't take place.
This radically
complicates the theoretical description of the nucleation kinetics
and  this case was completely investigated here.
The power-like dependencies with decreasing absolute value of
derivative on time were chosen as the model  laws of embryos
growth.
All main
characteristics of the nucleation period including the spectrum
of sizes of  the new phase objects  were obtained.  An analytical
approximate description of nucleation in different external
conditions was fulfilled.
The errors of such description are found and proved to be small.
\end{abstract}

\maketitle

\section{Introduction}

Kinetic description of a system under the  phase
transition is typical namely for the phase transition of the first
order.
The driving force of the first order phase transition
 implies the existence of metastability of a mother
phase. Namely this metastability can be regarded as a "source" of
the substance transformation from an unstable "mother" phase to a
stable "new" phase. Thus, kinetic investigation of the phase
transformation is rather actual for the phase transition of the
first order. For example, such an important characteristic as the
number of objects of a new phase can be calculated only on the
base of kinetic consideration.

Traditionally the physical model situation for the first order phase
transition was the case of the supersaturated vapor condensation
into the phase of liquid droplets. It was chosen as the base for
theoretical and experimental investigation due to the rather simple
description of the bulk phase. Namely in this case the
classical theory of nucleation was created in \cite{class}. This
gave the base for construction of the kinetic description of the
global evolution of the system under the first order phase
transition which was done in \cite{PhysRev94}, \cite{tmf}.
That's why we shall borrow the
terminology from the case of condensation \cite{PhysRev94}, \cite{PhysicaA95},
For example, the
supercritical embryos of a new phase will be called as droplets,
the process of the phase transformation will be called as
condensation, the period of formation of the main quantity of the
droplets will be called as nucleation, etc.

One of the
main features which allows to fulfill analytical constructions in
\cite{PhysRev94}, \cite{tmf} was the property of the avalanche
consumption of the mother phase by droplets. This means that the
quantity of the mother phase molecules absorbed by a small droplet
is many times smaller than the number of molecules absorbed during
the same time by a droplet of a big size.

In a free
molecular regime of the substance exchange one can write that
absorption coefficient $W^+$ is proportional to the surface of
droplet, i.e. to the square of the droplet radius $R$. Then $W^+$
grows in time $t$ under the constant power of metastability as $W^+
\sim t^2$ and the number of molecules $\nu$ accumulated by a
droplet grows as $t^3$. The big power $3$ allows to apply in
\cite{PhysRev94}, \cite{tmf} some effective methods which can be
reduced to  the first steps in some iteration procedures, formulated
there.

Under the diffusion regime one can get $\nu \sim t^{3/2}$
and even this not so rapid growth is also sufficient for the
application of the methods from \cite{PhysRev94} based on the first
iterations.

All possible regimes of the substance consumption lies between the
free molecular regime and the diffusion regime and namely this
property allows to establish in \cite{tmf} the general feature -
the avalanche consumption of metastable phase. This property isn't
only the key to get the numerical results but allows to combine
the collective regime of the metastable phase consumption and the
existence of the
density profiles in
a unique  approach
\cite{PhysicaA95}, \cite{tmf}. This possibility plays the fundamental
role in the justification of approaches formulated in
\cite{PhysicaA95}.

But sometimes the rate of droplets growth decreases in time. This
property doesn't allow to apply iteration methods from
\cite{PhysRev94}, \cite{tmf}. This situation requires  a special
consideration which will be given below. The situation with
decreasing rate of growth  takes place rather often in the
processes of cementation, structural transformations,
morphological transitions, etc.

Rigorously speaking one has to state that the avalanche
consumption is the characteristic feature of phase transition, the
justification is based on the geometrical features (i.e. on the
dimensionality of space). So, the situation with the decreasing rate
of growth isn't characteristic for the pure transition in a  rigorous
sense. But in practice the decreasing rate of growth is also
considered in frames of the first order phase transition in
an
ordinary sense. So, it is worth doing here through
a presentation of
the analytical theory of the process.

We shall consider two types of external conditions which seems to
be
typical and are widely spread both in theoretical and practical investigations.
The
first type is a situation of decay. It
takes place when the power of metastability
is attained in the system rather rapidly and later there is
absolutely no external influence upon the system.
The power of
metastability is fully determined here by the condensation
processes.

The second type typical external conditions are conditions of dynamic
type. Really, ordinary practically all
external conditions have rather smooth
behaviour in time. It means that the metastability is created
gradually. Formation of droplets and mainly the process
of growth consume vapor and
the power of metastability begins to fall. As the result the
process nucleation stops and one have the formed spectrum of sizes
of droplets with rather simple  law of evolution.

Now we shall discuss the law of the droplets growth.
Here we shall use the power-like dependencies of the droplets growth.
The power-like
dependencies are explained by the absence of some characteristic
dimensional scale like  in the scaling theory for the
second order phase transitions. The diffusion regime and the free
molecular regime have  also the power-like laws of growth. So, we
shall also take the power-like regimes here. Namely
$$
\nu =t^{\alpha}
$$
with parameter $\alpha$.
We shall call the power dependencies with  $0<\alpha < 1$ as the
slow rates of growth.

We shall use all results and definitions formulated in
\cite{PhysicaA95}, \cite{PhysRev94}, \cite{tmf}.
But, certainly the methods used in
these papers have to be reconsidered. They can not be directly
applied under the slow rate of growth and a new
methods
have to be developed.

We investigate the case of homogeneous phase transition. The
heterogeneous case can be investigated analogously to
\cite{PhysRev94}. In \cite{PhysRev94} it is shown that the homogeneous case
can be effectively
applied as the base for the consideration of heterogeneous case.
The same takes place here also.

\section{Evolution equations}

The situation with the slow rate of growth is more complex
because now not only the droplets with relatively big sizes are
the main consumers of vapor. Here all droplets
takes place in the consumption of
vapor. But there is one simplifying  feature
which states  that now there is no need to consider the density
profiles or  profiles of metastability \cite{PhysicaA95},
\cite{tmf}. Really the
characteristic scale of the diffusion blurring is
$$
R_{diff} \sim \sqrt{t-t'}
$$
where $t$ is the current moment of time, $t'$ is the time of the
perturbation appearance. The same is valid for the thermal blurring.
So, when $\alpha< 3/2$ the diffusion blurring destroys the density
profile and we come to the collective regime of vapor consumption
(see \cite{PhysicaA95}, \cite{tmf}).

The procedure of derivation of the evolution equation is
absolutely the same as in \cite{PhysRev94}, one has only to substitute
the power $3$ or $3/2$ by the new power $\alpha$.
The statement about the leading role of the supercritical embryos
in the substance balance used in \cite{PhysRev94} is so strong,
that it
remains valid until $\alpha \rightarrow
0$. But the case $\alpha \rightarrow 0$ can be investigated
explicitly without this assumption which will be done separately
(due to the simple conservation laws)
below.
The same is valid for the statement about the applicability of the
quasistationary approximation for the rate of nucleation (the rate
of appearance of droplets).

The power of metastability will be characterized by the value of
supersaturation
$$
\zeta = (n-n_{\infty})/n_{\infty}
$$
where $n$ is the molecular number density of the mother phase,
$n_{\infty}$ is the
same value for the saturated  mother phase.

The evolution equation for the situation of decay can be written
as
$$
G = A \int_0^z (z-x)^\alpha \exp( - G) dx
$$
with a positive parameter $A$, which has the sense of the spectrum
amplitude. The spectrum looks like
$$
f \sim \exp(-G)
$$

In the situation of dynamic conditions we can
linearize the power of metastability
in the absence of nucleation as
we have done in \cite{PhysRev94}. Then we have
$$
G = A \int_{-\infty}^z (z-x)^\alpha \exp(bx - G) dx
$$
with the additional parameter of the mentioned linearization $b$.
Here
$$
f \sim \exp(bx-G)
$$

In the derivation of these equations we have used the following approximation
for the stationary rate of nucleation
\begin{equation} \label{expap}
f_s(\zeta) =
f_s(\zeta_0) \exp(D  (\zeta - \zeta_0))
\end{equation}
as  a function of the supersaturation $\zeta$ where index $0$
characterizes the base of decomposition. Here $D$ is the
derivative of the free energy of  a critical embryo formation
over $\zeta$.

This approximation isn't something special but simply the
Klapeiron-Klausius equation. Ordinary this equation is used for the
dependence of $n_{\infty}$ over temperature but there is no
difference to what characteristic we shall apply this procedure of
approximate derivation.
We have to mention that  renormalizations to describe
the nucleation kinetics in non isothermal conditions are based on
the Klapeiron-Klausius equation for the density of the saturated
vapor.
Here the analogous renormalizations in order to take into account
the non isothermal effects can be also done.

\section{Decay of metastable state}

At first we shall investigate the  evolution of the system under
external conditions of the decay type.
Certainly, one can act in the manner analogous to the
approach of the  universal solution \cite{tmf} for the case of
the avalanche consumption of the mother phase.  After the
evident rescaling $z \rightarrow A^{1/4} z, x \rightarrow
A^{1/4}x$ one can come to
$$
G =  \int_0^z (z-x)^\alpha \exp( - G) dx
$$
with no parameters. Then  $G$ is the universal function and the
zero momentum
$M_0 = \int_0^{\infty}$ is the universal constant. Since $M_0$ is
the coefficient in the asymptotically leading term
$M_0 z^\alpha$ in expression for $G$ after the end of nucleation
we have got all  information necessary for the further
evolution. This solves the problem of the nucleation description.

Having determined the duration of the nucleation period $\delta z$
as
$$
N(\delta z) = N(\infty) (1-\epsilon)$$
with some small $\epsilon$ and $N(z)$ as the number of  droplets
appeared before $z$ one can get for $\delta z$ some fixed
constant. The value $N(z) \sim \int_0^z \exp(-G) dx$ is also the
universal function and $\delta z$ as the root of equation with no
parameters will be the universal function.

Now we shall determine the form of the size spectrum of droplets
explicitly.
The iteration procedure 
isn't effective because the initial approximation is too rough.
To  propose another iteration procedure we shall establish the new
property of the droplets growth typical for the regime of th e
slow growth.

{\bf The property  of effective size of growth}

Let us consider the law of the the vapor consumption by a separate
droplet. It is shown in fig. 1. For every moment $z$ (or $ t(z)$)
we can find the boundary $z_0$ with two properties
\begin{itemize}
\item
$z_0 \ll z$
\item
$\nu(z) -  \nu(z_0) \ll \nu(z)$

\end{itemize}

One can see that for
$z_0 = \alpha z /p , \ \ \ p \sim 2 \div 3
$
both properties
are satisfied.

It means that the qualitative picture is the following: rather
soon the droplet accumulates the main quantity of vapor and during
the rest of the evolution time the accumulation isn't too
important.
Certainly, this moment of accumulation grows in time.

This picture  allows to suggest the following initial
approximation:

{\it
For every $z$ one can imagine that practically immediately the
number of molecules of a
given
droplet (with coordinate $z$) attains the value
$\nu(z_0)$ and then the droplet doesn't grow.
}

One has to stress that for different $z$ the values $z_0$ are also
different.

This property allows to use the approximation with $\alpha = 0$
as initial approximation, which has an evident advantage.

Here one can see the  property  of a
real collective consumption of vapor -

{\it
Practically all droplets (except small ones with sizes less than
$z_0$)  consume vapor in approximate equal quantities. This is
really the collective consumption of vapor (Earlier this term was
used simply for vapor consumption in a fixed point by many
droplets in different quantities. Now droplets consume vapor in
equal quantities and
we
shall speak about the equal collective
consumption.)
}

The case $\alpha = 0$ can be solved
analytically, because it can be reduced to
the ordinary differential equation of  the first order.
This leads to
$$
\exp(-G) = \frac{1}{1+z}
$$

One has to note that the last equation leads to the infinite
number of droplets appeared in the process of nucleation. This  is
an error.
This error is initiated by the inapplicability of approximation
(\ref{expap}) at small supersaturations. Really, at
$G \sim n -n_{\infty}$ the rate of nucleation has to be  vanished,
but approximation (\ref{expap}) gives the finite value of the
nucleation rate at any $G$.

Nevertheless we aren't interested at the long tail $\sim z^{-1}$ of the
size spectrum because this tail can not be directly seen in
experiment and all integral values can be explicitly calculated.
Really, the total number of droplets can be easily obtained, it is
convenient to do even in initial variables.

The case $\alpha = 0$ corresponds to a fixed number $\nu_{fin}$ of
a number of molecules inside the droplet. This quantity doesn't
depend on time, in rescaled units $\nu_{fin} = 1$.
The total number of droplets is
$$
N_{total} =
\frac{n -n_{\infty}}{\nu_{fin}}
$$

Formally we can choose the cut-off of the spectrum at the size
where $N(z)$ equals to $N_{total}$.

{\bf Iteration procedure}

One can see that even in the case of small positive
$\alpha$ the spectrum of
sizes  is well localized which allows to use the
iteration procedure
$$
G_{i+1} =  \int_0^z (z-x)^\alpha \exp( - G_i) dx
$$
Here the lower index denotes the number of iteration. As initial
approximation we choose here the analytical solution for
$\alpha = 0$ , i.e.
$$
G_0 = - \ln(x+1)
$$
This radically differs from consideration of the situations with
avalanche regimes of growth, where $G_0 = 0$.

The first iteration can be easily calculated. We have
$$
G_1 =
\int_0^z (z-x)^{\alpha} \frac{1}{1+x} dx
$$
which can be easily calculated
$$
G_1 =  - \frac{z^{\alpha+1}}{z+1} \Phi(- \frac{z}{z+1}, 1,
\alpha+1)
$$
where
$$
\Phi(z,s,v) = \sum_{k=0}^{\infty}
\frac{z^k}{(v+k)^s}
$$
is the
standard special function \cite{Rujik}.

The spectrum based on $G_1$ , i.e.
$\exp(-G_1)$ can be considered as a rather precise approximation.
It can be shown analytically and numerically. Namely, the relative
error in the number of droplets, i.e. in
$$
N(z) = \int_0^z \exp(-G(x)) dx
$$
is less than $0.05$ for $\alpha < 0.5$. The values $\alpha$ from
$1/2$ up to $3/2$ can be considered analogously on the base of
iterations started from solution with $\alpha = 1$. This solution
can be also found analytically since the evolution equation can be
reduced to the second order ordinary differential equation with no
explicit dependence on the argument, which can be easily integrated.
Then is is necessary only to take one iteration step and then to
integrate $\exp(-G_1)$ over time to get the number of droplets.

Figure 2 shows the relative error in the total number of droplets
as a function of $\alpha$. The integration is fulfilled up to
$z=10$ which is absolutely sufficient for any case except $\alpha =
0$ and in this case the error is zero (certainly with artificial
cut-off the same will be numerically).
The deviation from zero at small $\alpha$ is initiated only by
a finite step of integration.
One can see that the relative error is rather small. This can be
shown not only numerically but also analytically.

Figure 3 shows the characteristic forms of spectrums. Two cases
are shown - $\alpha = 0.2$ and $\alpha = 1$. One can see two pairs of curves
coinciding at the beginning and approximately at the end.
Every pair corresponds to one case and presents the precise solution and
our approximation. The case $\alpha = 1$
has analytical solution and it is out of consideration here but
one can see that even in this case our approximation works good.
In this case the back side of spectrum is more sharp than for
$\alpha = 0.2$.
One can also notice the relative invariancy of the form of
spectrum to the concrete choice of $\alpha$ for small $\alpha$.

One has to mention that the consumption of vapor by all droplets
requires to reconsider slightly the characteristic size of
consumers in the statements about the quasistationarity of the
nucleation rate and about the size of the main consumers of vapor
\cite{PhysicaA95}. But even with these modifications both
statements can be proven (and under dynamic conditions also)

\section{Dynamic conditions}

The simple rescaling $z \rightarrow z A^{1/\alpha}$,
$x \rightarrow x A^{1/\alpha}$ brings evolution equation to
$$
G =  \int_{-\infty}^z (z-x)^\alpha \exp(bx - G) dx
$$
and allows to cancel parameter $A$.  But it isn't too simple to
cancel parameter $b$ because earlier \cite{tmf}
the condition, which
allows to cancel this parameter was the following:
The point of decompositions has to be chosen as the maximum of
supersaturation (or as the maximum of $f$)

In situation with $\alpha=0$ one can not satisfy this condition.
Moreover, $f$ increases in time. The analytical solution
in the case $\alpha = 0$ is the
following
\begin{equation}\label{w}
f = \exp(bx-\ln(\frac{\exp(bx)}{b}+1)) \equiv f_0
\end{equation}
This solution can be got after differentiation of the evolution
equation which brings it to
$$
\frac{dG}{dx} = \exp(bx-G(x))
$$
One can easily integrate this equation
$$
\int \exp(G) dG = \int \exp(bx) dx
$$
and come to (\ref{w}).

When $\alpha \ne 0$ one can see the maximum of $f$ and when
$\alpha=0$  the explicit analytical solution has been presented.

For $\alpha > 0$ one can put condition on maximum.
Then for $\alpha \ne 0$ we have the universal solution depended on
$\alpha$. This solution satisfy the following
universal equation
$$
-\ln(f) + \alpha \int_{-\infty}^0 (z-x)^{\alpha-1} f(x) dx z
=  \int_{-\infty}^z (z-x)^\alpha f(x) dx
$$
the ideology here is absolutely analogous to the already
investigated situations.

{\bf Iteration steps }

Now we shall present the methods to get the
explicit form of the size spectrum. The main feature which has to
be taken into account is the long infinite tail of the spectrum
for $\alpha = 0$. The explicit form of spectrum in this case
for $b \sim 1$ is
drawn in figure 4. One can see that the curve is slightly blurred
because here an approximation which will be discussed later is
also drawn.

The most interesting region here is the region of the rapid
increasing of $f$. If we choose $b=1$ we get this region at $x
\sim 0$.
Later we shall see that it
is  reasonable to put $b \sim 1 $ for all $\alpha$.

One can formulate an
important feature: Certainly, for $\alpha>0$ the
spectrum $f$ lies lower than the  spectrum $f_0$ in the case
$\alpha = 0$. Having used the property of effective size of growth
 one can come to conclusion that the maximum
of $f$ lies near $x=0$ when $b=1$. So, then one can put $b=1$ for
all cases with small $\alpha$.

We shall use $f_0$ as initial  approximation in the iteration
procedure
$$
G_{i+1} =  \int_{-\infty}^z (z-x)^\alpha \exp(bx - G_{i}) dx
$$
$$
G_0 = - \ln(f_0) + b x
$$
The first iteration can be presented as
$$
G_1 = \int_{-\infty}^z (z-x)^{\alpha}
(\frac{\exp(bx)}{b} +1 )^{-1} \exp(bx) dx
$$
To calculate $G_1$ one can act in two manners. The first one is to
invent approximation for $(z-x)^{\alpha}$. as we have already
mentioned this function has behaviour drawn in figure 1. We see
that that it can be well approximated by
a straight line as it is drawn in figure 5.
This approximation allows the analytical calculation of $G_1$
since the integral can be reduced to
$$
\int \frac{1}{(x+c_1)^2-c_2} dx
$$
and to
$$
\int \frac{\ln(x+c_3)}{(x+c_4)^2-c_5} dx =
= \frac{1}{2c_5}\arctan(\frac{c_5 x}{c_4^2 +c_5^2 +c_4 x}
\ln[(c_3 - c_4)^2  +c_5^2] -
$$
$$
- \frac{1}{2c_5} Cl_2(2 \Theta + 2 \Phi ) +
\frac{1}{2 c_5} Cl_2  (2 \Theta_0 + 2 \Phi ) -
\frac{1}{2c_5} Cl_2 (\pi - 2 \Theta) + \frac{1}{2 c} Cl_2 (\pi - 2
\Theta_0)
$$
Here
$$
\tan \Theta = \frac{x+c_4}{c_5}
$$
$$
\tan \Theta_0 = \frac{c_4}{c_5}
$$
$$
\tan \Phi = \frac{c_3 - c_4}{c_5}
$$
and
$$
Cl_2(z) = - \int_0^z \ln(2 \sin (\frac{x}{2}) dx
$$
is the Klausen integral which is the standard special function.

The second method which suggests an approximation for $f_0$
seems to be more attractive.
This approximation is the following
$$
f_0 \equiv \frac{1}{1+\exp(-x)} \approx f_{ap}
$$
where
$$
f_{ap} = \exp(x) - \exp(2x) + \frac{1}{2} \exp(3 x)
$$
for $x<0$
and
$$
f_{ap} = 1 - \exp(-x) + \frac{1}{2}\exp(-2x)
$$
for $x>0$.

Namely this approximation together with precise solution is drawn in
figure 4. Certainly at $x=0$ there is no difference between
precise result and approximate one.
With the help of this approximation all integral terms can be
calculated in terms of Gamma-function.

When the first approximation $G_1$ is calculated we see that the
specrtum $f_1  = \exp(x - G_1) $ in the first approximation has
one maximum $f_{1\ max}$ at $x_{1\ max}$
and at the region $x<x_{1\ max}$ the spectrum $f_1$
approximates the real
spectrum quite satisfactory.
Moreover we have $x_{max} \approx x_{1\ max}$,
where $x_{max}$ is the coordinate of real maximum and
$f_{max} \approx f_{1\ max}$ where $f_{max}$
is the amplitude of a real spectrum. These facts can be
proven analytically.
The form of size spectrums for characteristic values of $\alpha$
are drawn in figure 6.

One can see one important feature here. The spectrum in the first
iteration rapidly turns to zero while the real spectrum has very
long and smooth tail. This occurs because the amplitude of the spectrum
(and also of the first iteration) is
seriously less than
the amplitude of the
solution for $\alpha = 0$. Now we can correct this error.
When $\alpha$ is growing then $x_{1\ max} $ goes away from
$x_{max}$;  $f_{1\ max}$ goes away form $f_{max}$ and $f_0(x_{1\
max})$ goes away from $f_{1\ max}$. But even when $\alpha = 0.5$
(this is the realistic boundary of our constructions because for
bigger $\alpha$ one can use decompositions starting from $\alpha =
1$) the difference between $f_0(x_{1\
max})$ and $f_{1\ max}$ is small. This allows to suggest the
cut-off of the zero approximation
$f_{0\ cut}$ which is
$$
f_{0\ cut} = f_0
$$
for
$x<x_{1\ max}$
and
$$
f_{0 \ cut} = f_0 (x_{1\ max}) =  f_{1\ max}
$$
for
$x>x_{1\ max}$.

On the base of this initial approximation one can reproduce all
constructions and get the spectrum $\hat{f}_1$ in
the first approximation, which lies higher
than the "previous first approximation". It is drawn in figure 7
where this curve lies higher then the previous first iteration.

One can analytically show that the first advanced iteration
describes the form of the hill  (the head of the spectrum)
satisfactory. But there remains the
long tail and it is absolutely impossible to describe t on the base
of the iteration method because the duration of the back tail can
be infinitely long. Then one can propose come asymptotic methods
to solve this problem.

{\bf Asymptotics }

We shall construct solution for $x \gg 1 $. At first we can see
that it is possible to write approximately evolution equation as
$$
G(z) = \int_0^z (z-x)^{\alpha} exp(x-G(x)) dx
$$
instead of
$$
G(z) = \int_{-\infty}^z (z-x)^{\alpha} exp(x-G(x)) dx
$$

Now we shall use the property of effective size of growth.
Since the amplitude of
spectrum doesn't vary too rapid during this time one can take away
the size $(z-x)^{\alpha}$ out of the integral and speak about the
mean size $\bar{\rho}$. Then
$$
G(z) = \bar{\rho} \int_0^z  exp(x-G(x)) dx
$$
For $\bar{\rho}$ one gets the following expression
$$
\bar{\rho} = \frac{
\int_0^z (z-x)^{\alpha} dx
}{\int_0^z  dx
} = \frac{z^{\alpha}}{\alpha+1}
\approx
z^{\alpha}
$$

Having differentiated $G$ over $z$ one has to notice that
$\bar{\rho} = z^{\alpha}$ is rather slow function and there is no
need to differentiate it. Then
$$
\frac{d G}{dt} =
z^{\alpha}\frac{d}{dz} \int_0^z \exp(x-G(x)) dx=
z^{\alpha} \exp(z-G(z))
$$

After intergation we have
$$
\int \exp(G) dG = \bar{\rho} \int \exp(x) dx +const
$$
and
$$
G = \ln(1+ \bar{\rho}  \exp(z))
$$

This solution has asymptotics
$$
G \rightarrow \ln( \bar{\rho}) + z
$$

This asymptotics leads to the following expression for the size
spectrum
$$
f \sim \exp(x - G) = exp(z-\ln( \bar{\rho})-z) =( \bar{\rho})^{-1}
$$
and finally
$$
f \sim \frac{1}{z^{\alpha}}
$$

Here one can introduce the arbitrary shift $\gamma$ and the
arbitrary amplitude $\beta$. Then
$$
f \sim \frac{\beta}{(z-\gamma)^{\alpha}}
$$
These parameters can be determined by requirement of the smoothness
of spectrum and its
 derivative in the point of transition from the advanced first
 iteration to the asymptotics.  Were shall we choose this point?

At $z=0$ the spectrum begins to grow, at $z=z_{max}$ it attains
the maximum, it is reasonable to imagine the hill as symmetric one
an dto say that at $z =  2 z_{max}$ the hill is over. Namely
at this point we can speak about the beginning of the asymptotic.

The conditions
$$
\frac{\hat{f}_1}{dz} = - \alpha
\frac{\beta}{(z-\gamma)^{\alpha+1}}
$$
and
$$
\hat{f}_1 =\frac{\beta}{(z-\gamma)^{\alpha}}
$$
at $z =  2 z_{max}$
give the following expressions
$$
\gamma = 2 z_{max} + \alpha \hat{f}_1(z_{max}) /(d\hat{f}_1 /
dz)|_{z=z_{max}}
$$
$$
\beta = \hat{f}_1(z_{max}) (z-\gamma)^{\alpha}
$$

The characteristic forms of spectrums and analytical
approximations are drawn in figure 8.
The letters A-E  denote the pairs of curves corresponding to
different $\alpha$  from $0.1$ to $0.5$ with a step $0.1$. The
capital letters mark the precise numerical solutions, they lies
near approximate solutions.

The accuracy of the theory can be estimated by the error in the
droplets number
$$
\epsilon = \frac{|N-N_{ap}|}{N}$$
 is drawn.
Here $N$ is precise value found from numerical solution and $N_{ap}$
is the number of droplets found from the presented approximate solution.

In figure 9 the maximum
of $\epsilon$ over $z$  is
drawn as a function of $\alpha$. It can be seen that it is rather
small. Here the maximum value of $z$ is chosen as $30$.

Now we shall analyze the accuracy of calculations. In figure 10
the relative error for $N(z)$ at some final value is drawn.
Here we consider the "final" values for $N$. Certainly, the terms "the final
values" are illegal, because it is clear that for $\alpha <1$ the
finite number of droplets can not satisfy the balance of
substance.   So, we need to throw away th e tail, which is thin
but still infinite (it contains infinite number of droplets).

In  consideration of extremely long
asymptotic tails
we have to take care about the accurate behavior of
asymptotics. The combination of iteration solution and asymptotics has
to be performed here in  a slightly
other special point. We shall mark this point
as $z_{bound}$.

To get $z_{bound}$  we shall fix the beginning of
nucleation more precise. On the base of iterations we get
$f_{max}$. Then we get the value $f_{st}$ of the amplitude at  the
beginning of nucleation as $f_{st} = f_{max} / \exp(1)$. Then we
can get the coordinate $z_{st}$ of the beginning of nucleation as
$z_{st} = \ln(f_{st})$.
Then
$z_{bound}$ will be calculated according to the old recipe but
with
a new time of the beginning of nucleation
$$
z_{bound} = z_{st} + 2 (z_{max} - z-{st}
$$

In figure 11 the values of the droplets number are compared at a
"final" value $z_{fin}$. Two values $z_{fin} =50$ and $z_{fin}=
70$ are chosen here.
The value of $\epsilon$ is drawn.
 One can see three broken lines. The shapes of
two of them
are approximately the same at small $\alpha$
 - these lines are the relative
errors in the droplets number for different $z_{fin}$. It is clear
that the dependence on $z_{fin}$ at such values is very smooth and
we practically attains the limit case (precisely speaking, it can
not be done). The step in calculations $dx$ was chosen as
$dx = 0.05$. The third line  at small $\alpha$
lies below and has  a more smooth shape. It is the same
value $\epsilon$ for $z_{fin} = 50$ but
calculated with a  step $dx=0.025$. It isn't too far from the two
mentioned curves. So, the necessary accuracy is attained.

Now one can analyze the
behavior of the error. The decrease of the error at $\alpha
\approx 0.7$ is caused simply by compensation of different sources
of errors, the value $|N-N_{ap}|$ changes
a sign here. More
interesting fact is the reduction of the error at $\alpha \sim 1$. The
reasons are the following:
Every spectrum has the tail and the head. The head is rather short
and it is localized well and can be described with the help of
iteration method. The tail has to described by asymptotics.
The main source of error is th absence of account of the
influence  of surplus substance appeared in the head of spectrum
in evolution at asymptotics. When we are going to extremely long
tails this influence will disappear.

 When
$\alpha$ attains $1$ the quantity of the droplets in tail
at intermediate $z$ is
small in comparison with the quantity of droplets in the head. So,
it is reasonable to take in to account only the head with the help
of iterations (Certainly, now the number of droplets wil be counted
on the base of iterations not only until $z_{bound}$ but also fo
r$z>z_{bound}$).
So, now
$$
N_{ap} = \int_{-\infty}^{\infty} \tilde{f}_1(x) dx
$$
$$
\tilde{f}_{1} = \exp(x-\tilde{g}_1)
$$
$$
\tilde{g}_1 = \int_{-\infty}^z (z-x)^{\alpha} \hat{f}_0 (x)
$$
$$
\hat{f}_0 = max f_1
$$
for  $x>x_{max\ 1}$
and
$$
\hat{f}_0 =  f_1
$$
for  $x<x_{max\ 1}$.
Here
$max f_1$
is the maximal value of $f_1$ and $x_{max\ 1}$ is the
corresponding argument. To get these values it is necessary to
calculate
$$
f_1 = \exp(x-g_1)
$$
$$
g_1 = = \int_{-\infty}^z (z-x)^{\alpha} f_0 (x)
$$
where $f_0$ is the analytical solution corresponding to $\alpha =
0$.

The result is drawn in figure 12. Here we draw the relative error
$\epsilon$ at $z=50$. The step of calculations was $dz= 0.05$. We
see that even for $\alpha = 3/2$ the result is good even without
asymptotics and even with initial approximation corresponding to
$\alpha = 0$. For $\alpha > 3/2$ the result was presented in
\cite{tmf}. All situations are solved now.

We have to stress that initial approximation as the analytical
solution for the case $\alpha = 1$ is better
than initial approximation corresponding to $\alpha = 0$.
Then the accuracy
will be also better. Now we shall present the analytical solution
for $\alpha =1$.
The evolution equation is
$$
g = \int_{-\infty}^{z} (z-x) \exp(x-g(x)) dx
$$
For $\phi = -x+g$ this equation is
$$
\phi = \int_{-\infty}^{z} (z-x) \exp(-\phi(x)) dx
$$
Having differentiated two times we get
$$
\frac{d^2 \phi}{dz^2}
=
\exp(-\phi)
$$
This second order differential equation doesn't contain the
argument explicitly, which allows to integrate it.
Let $\phi$ be the argument $u$, $d\phi / dz$ be the function $y$.
Then  $d^2 \phi / d z^2 = y dy / du$. Integration gives
$$ -
\frac{y^2}{2} =
 \exp(u) + c_1
 $$
 From the boundary conditions we get
 $$
 c_1 = - \frac{1}{2}
 $$
 Then
 $$
 d \phi / dz = \sqrt{c_1 - 2 \exp(\phi)}
$$
Integration gives
$$
\int \frac{d \phi}{\sqrt{c_1 - 2 \exp(\phi)}}
= x +c_2
$$
This integral can be easily taken analytically  which
gives the analytical  expression for the spectrum in this case.
The constant $c_2$ can be got from the limit behavior $\phi
\rightarrow
- x $ when $x \rightarrow -\infty$.

This solution is very fruitful for description of
situations with $\alpha$ close to $1$. Results are given in figure
13 which is analogous to the figure 12 but with another initial
approximation. One can see here the curve A, which demonstrates the
error of the first iteration in direct iteration method.
Here
$$
N_{ap} = \int_{-\infty}^{\infty} f_1(x) dx
$$
$$
f_1 = \exp(x-g_1)
$$
$$
g_1 = = \int_{-\infty}^z (z-x)^{\alpha} f_0 (x)
$$
and $f_0$ is the analytical solution corresponding to $\alpha =
1$.

 We see that the
error is even greater than the error in figure 12. It is so because
simply the size (the amplitude,
not the shape) seriously diminishes. This is
the same as
we have seen for smaller powers. This diminishing is the main
source of error.
So, we need to reexamine the initial approximation.

The first way is to act in a style like it has been done with the
advanced iterations quite above. Here
$$
N_{ap} = \int_{-\infty}^{\infty} \tilde{f}_1(x) dx
$$
$$
\tilde{f}_{1} = exp(x-\tilde{g}_1)
$$
$$
\tilde{g}_1 = \int_{-\infty}^z (z-x)^{\alpha} \hat{f}_0 (x)
$$
$$
\hat{f}_0 = max f_1
$$
for  $x>x_{max\ 1}$
and
$$
\hat{f}_0 =  f_1
$$
for  $x<x_{max\ 1}$.
Here
$max f_1$
is the maximal value of $f_1$ and $x_{max\ 1}$ is the
corresponding argument. Then
$$
f_1 = \exp(x-g_1)
$$
$$
g_1 =  \int_{-\infty}^z (z-x)^{\alpha} f_0 (x)
$$
and $f_0$ is the analytical solution corresponding to $\alpha =
1$.
The results are given by the curve B.

We can act also  in another manner. We can rescale the number of
molecules in a liquid phase $g$ to have the approximate equal
amplitudes. One can calculate
constants
$$
q (\alpha)= \int \exp(-x) x^{\alpha} dx
$$
and instead of
$g$ consider $g/q$ (both for the current $\alpha$ and for $\alpha
= 1$. This corresponds to the approximate equal $g$ at $z=0$. The
error is drawn as the curve C. One can see that the error seriously
diminished.

One can also require the approximate equality in derivatives of
$g$ on $x$ at $t_*$
as it is considered in the balance for establishing of $t_*$
(see \cite{PhysRev94}). Then the constants $q$
will be
$$
q (\alpha)= \alpha \int \exp(-x) x^{\alpha-1} dx
$$
The result is drawn by the curve D. It is seen that the error is
practically the same. It corresponds to the property of  approximate
universality observed in \cite{tmf}. Now we see that this
approximate
universality goes also for the case of intermediate $1 < \alpha
< 3/2 $.

Here and above the values of $q$ were calculated on the
base of the ideal supersaturation, i.e. on the base of $\exp(x)$.
The results will be even better if we take $q$ calculated of the
base of solution at $\alpha = 1$.

One can not calculate here iterations analytically, but can act in a
manner presented in \cite{PhysRev94}.
Certainly, the leading term here will be $z N$ where  $z$ is the
coordinate of the maximum of peak of spectrum.

The situation of decay is much more simple and
we can use the standard
iteration solution given by formulas
$$
N_{ap} = \int_{0}^{\infty} f_1(x) dx
$$
$$
f_1 = \exp(-g_1)
$$
$$
g_1 = = \int_{0}^z (z-x)^{\alpha} f_0 (x)
$$
where $f_0$ is the analytical solution corresponding to $\alpha =
1$.
The error is drawn in figure 14. It is small. The slight
diminution when $\alpha$ grows is caused only by  numerical errors.

One has also to keep in mined that in situation with long tails
one has to take into account the higher derivatives of the free
energy of the critical embryo. This account is rather easy since
we know the general solution.
It can be done  both by iteration approach and by perturbation
technique.

Also one has  to  take care  about  the validity of linearization
of ideal supersaturation for the long tails. But it very simple to
correct the asymptotics for deviations from the linear case.

\pagebreak



 Figure 1

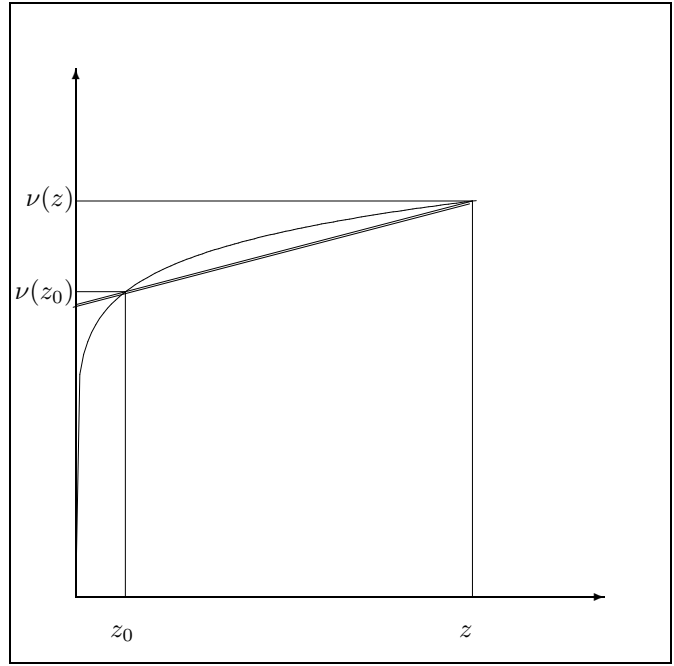
 Linear approximation of the size of growth

 Power $\alpha = 1/8$

  The bold line presents the approximation

\pagebreak
%

 Figure 3

 Forms of spectrums

\pagebreak

%

 Figure 4

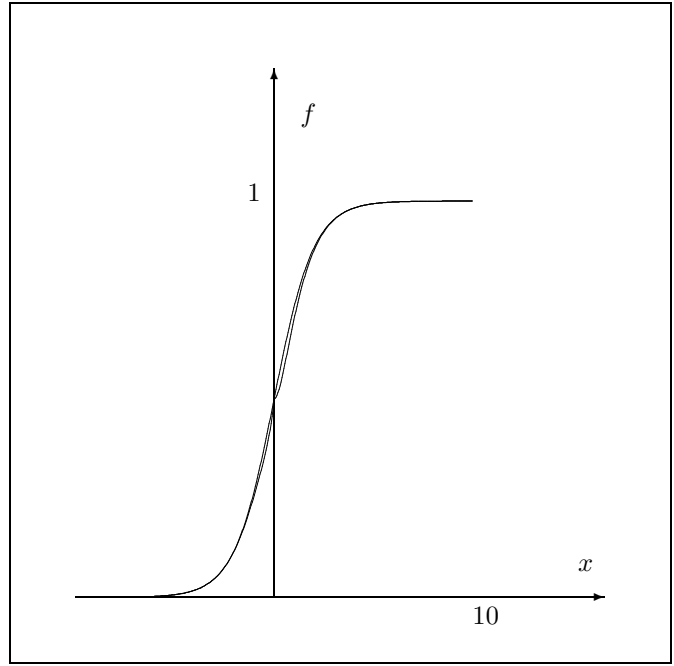
 Forms of spectrum and approximation

   for $\alpha = 0$

\pagebreak

%

 Figure 6

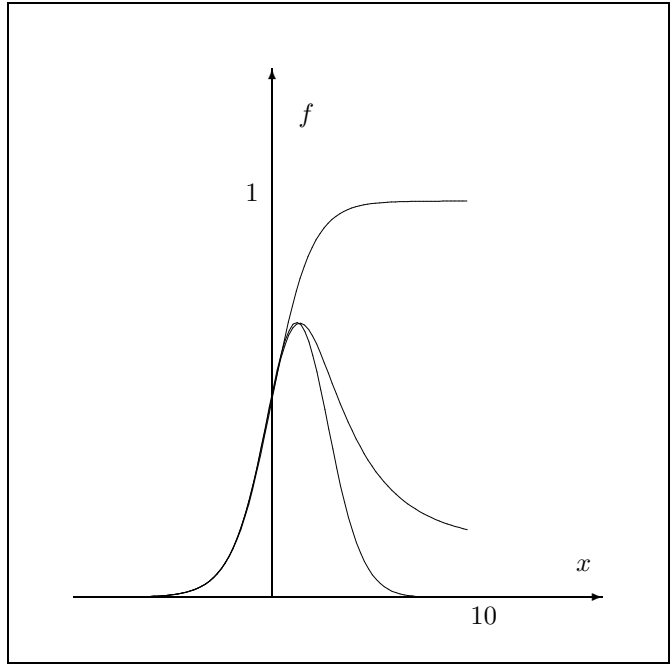
 Forms of spectrums in the first iteration

\pagebreak

%

 Figure 7

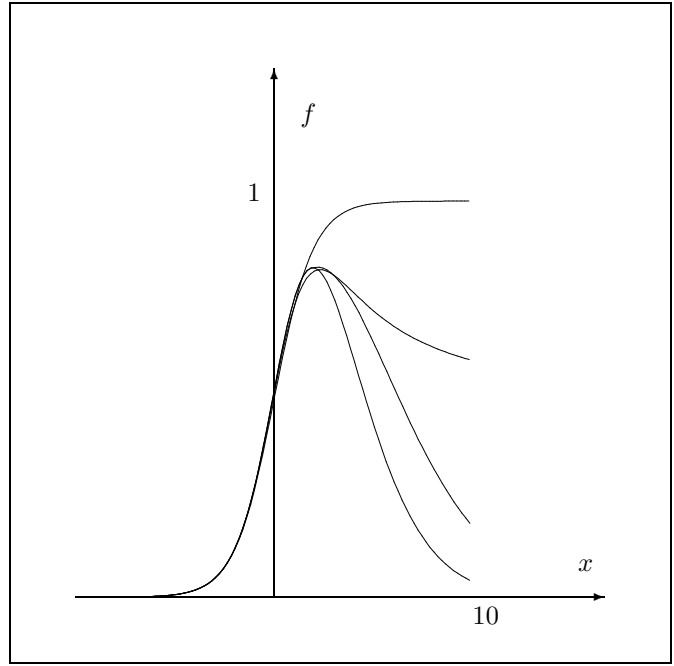
 Forms of spectrums in the first modified iteration

  $\alpha = 0.2$

\pagebreak

%

 Figure 8

 Forms of spectrums

\pagebreak

%

 Figure 3

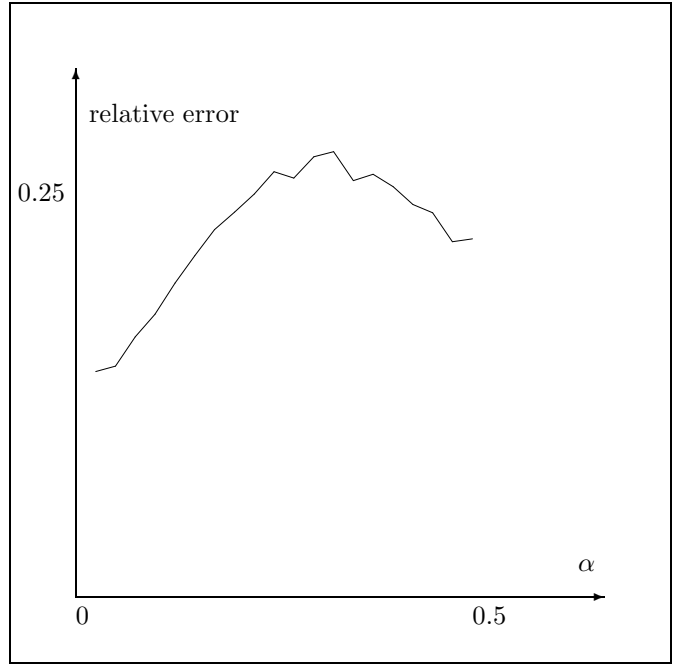
 Maximal relative errors

\pagebreak

%

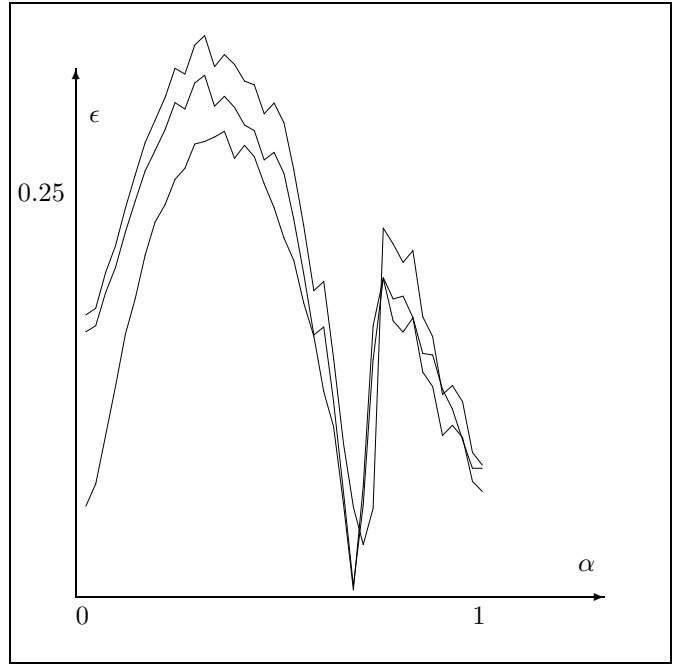
 Figure

 Relative errors

  The lenght is 70

  The step is 0.05

\pagebreak

%

 Figure

 Relative errors for pure iterations

  The lenght is 50

  The step is 0.05

\pagebreak

%

 Figure

 Relative errors for pure iterations

  The lenght is 50

  The step is 0.05

  Basic power 1

\pagebreak

%

 Figure

 Relative errors for pure iterations

  The lenght is 50

  The step is 0.05

  Basic power 1

  decay

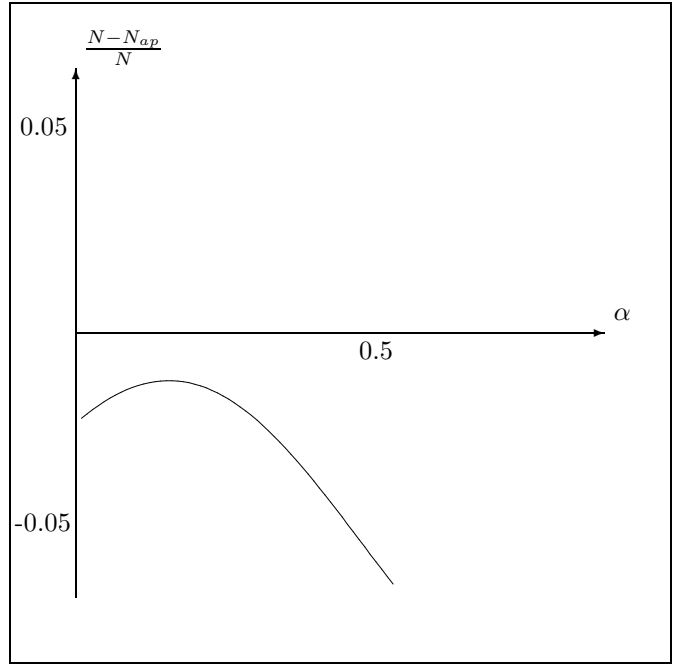
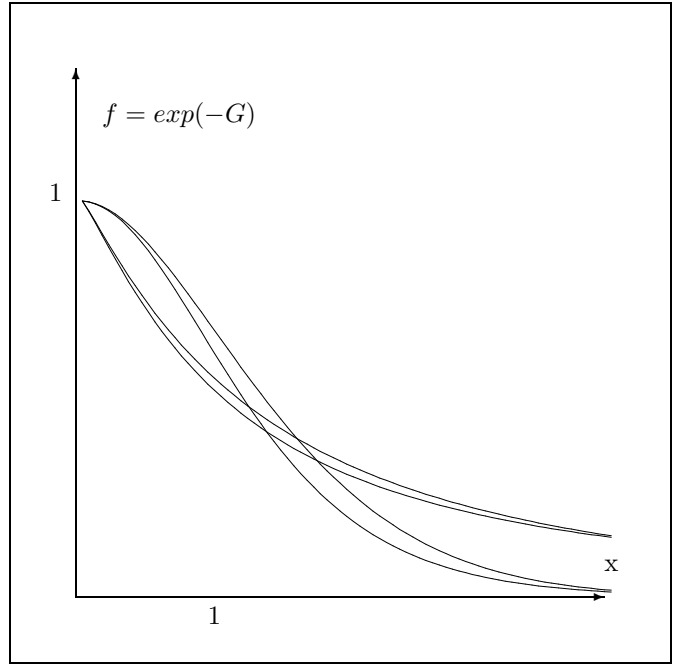
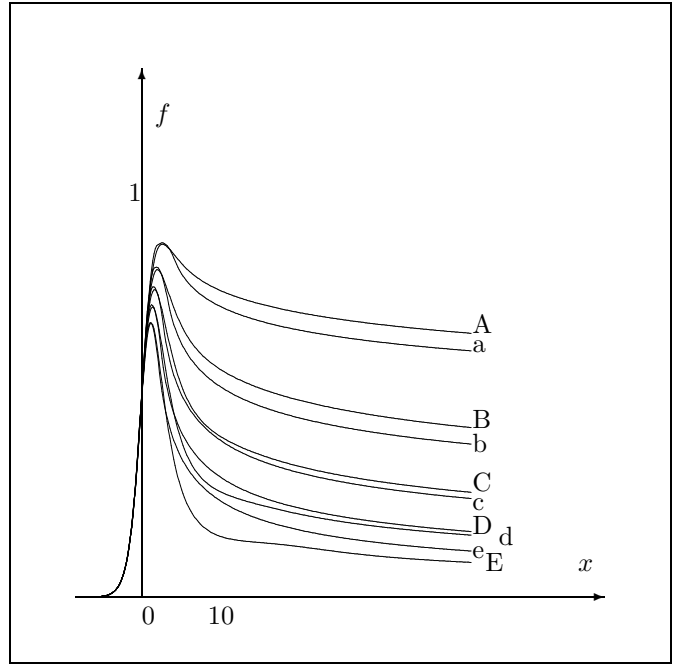
\end{document}